\documentclass[prd,aps,preprint,tightenlines,superscriptaddress]{revtex4}
\usepackage{mathrsfs}
\usepackage{amssymb}
\usepackage{color}
\usepackage{epsfig}
\usepackage{graphicx}
\usepackage{amsmath}

\begin{document}
\preprint{UMD-PP-09-033}

\title{R-Parity Breaking via Type II Seesaw, Decaying Gravitino Dark
Matter and PAMELA Positron Excess}

\author{Shao-Long Chen}
 \affiliation{Maryland Center for Fundamental Physics and Department of
Physics, University of
Maryland, College Park, Maryland 20742, USA }
 \author{Rabindra N. Mohapatra}
 \affiliation{Maryland Center for Fundamental Physics and Department of
Physics, University of
Maryland, College Park, Maryland 20742, USA }
\author{Shmuel Nussinov}
\affiliation{Tel Aviv University, Israel and Chapman College, California}
\author{Yue Zhang}
\affiliation{Center for High-Energy Physics and Institute of
Theoretical Physics, Peking University, Beijing 100871, China}
 \affiliation{Maryland Center for Fundamental Physics and Department of
Physics, University of Maryland, College Park, Maryland 20742, USA }

\date{\today}

 \begin{abstract}
We propose a new class of R-parity violating extension of
MSSM with type II seesaw mechanism for neutrino masses where an unstable
gravitino is the dark matter of the Universe. It decays predominantly
into three leptons final states, thereby providing a natural
explanation of the positron excess but no antiproton excess
in the PAMELA experiment. The model can explain neutrino masses without
invoking any high scale physics while keeping the pre-existing baryon
asymmetry of the universe in tact.
 \end{abstract}

\maketitle

\section{Introduction}
Recent observation of an excess of positrons in PAMELA
experiment~\cite{pamela}
in the energy range from 10 to 100 GeV has confirmed similar
observations several years ago by the HEAT~\cite{heat} and
AMS~\cite{ams} experiments. A great deal of discussion is currently
under way to understand this excess and its possible implications for
physics beyond the standard model. While it is quite possible that this
excess is of pure astrophysical origin~\cite{astro},
there is hope that this may be coming from the dark matter in our galaxy.
Many models have  been proposed to explain this excess in terms of
different kinds of dark matter.

Interpretation of the observations in terms of dark matter raises several
issues:
\begin{itemize}
\item  how to explain the lack of any excess in the hadrons?
 \item  how does one get an adequate enough positron production rate to
explain the excess?
\end{itemize}
These issues have been discussed in two broad classes of models for dark matter:
(i) stable dark matter pair annihilation~\cite{ann} or
(ii) decaying dark matter~\cite{decay}.

In the first category of models with
thermal production of dark matter relic density in the early universe,
the dark matter annihilation is constrained by the observed relic density
today. If one uses the same value for the annihilation cross section,
an additional enhancement is required to understand the observed positron
excess for popular dark matter density profiles,
e.g. Navarro-Frenk-White (NFW)~\cite{Navarro:1995iw} type.
There have been two ways to address this problem: (i) by increasing the local density of
dark matter and/or (ii) by introducing new hitherto
unobserved light particles~\cite{new} that can give rise to the so-called
Sommerfeld enhancement of cross sections at lower particle energies.

On the other hand,
in the alternative scenario involving decaying dark matter, there is no
correlation between the cross section that generates the relic density
and the magnitude of positron excess since the latter involves the decay
rate which is independent of the physics involved in relic density
generation. Therefore the second problem does not arise in this class of
models. For this reason, we focus on a decaying dark matter model in this
paper.

The issue of lack of hadrons can potentially be a problem
in both classes of dark matter models and requires further
model building with specific dark matter properties.

The decaying dark matter we consider is an unstable, long lived gravitino
which can arise naturally in supergravity models if gravitino is the
lightest
supersymmetric particle and if the model violates R-parity. The long
lifetime of gravitino required for it to be a viable dark matter is
automatically satisfied since the gravitino decay involves a
combination of the very weak gravitational
interactions as well as weak R-parity breaking~\cite{gravi}. Absence of
lepton number violation in any observed process to date justifies the second
assumption.

Let us start by discussing the familiar R-parity violating
models~\cite{bar}, i.e. the minimal supersymmetric standard model
(MSSM) with R-parity violating terms in the superpotential of type
$LLe^c$, $QLd^c$, $u^cd^cd^c$ and $LH_u$. In these models, one
generally expects both leptons as well as hadrons in the final
states of gravitino decay. Models of this type have been considered
in ref.~\cite{Ibarra:2008qg}. If however one kept only the $LLe^c$
term and drop all the others, then, the predominant decay mode of
the gravitino will only be to leptons as required to understand the
PAMELA data. However, one problem with this scenario is that if the
strength of this coupling (usually denoted by $\lambda$), is
larger than $10^{-7}$, this will erase the baryon asymmetry of the
universe. While the strength of this magnitude is also what is required
for understanding the PAMELA observations, it is too small to
explain neutrino masses via loop corrections, without assuming other
physics beyond MSSM, e.g. grand unification. Thus if we want a
minimal bottom-up approach to understand the PAMELA observation,
while at the same time keeping the baryon asymmetry of the universe
untouched and an explanation of neutrino masses with only TeV scale
physics, one must seek alternative models. This is what we do in
this paper.

We propose a new class of R-parity violating interactions that can
arise in extensions of MSSM which does three things using only
TeV scale physics: (i) it explains small neutrino masses and
mixings via the type II seesaw mechanism~\cite{type2}; (ii) it keeps the
baryon asymmetry of the universe untouched and (iii) it is
able to explain the leptophilic nature of the PAMELA observations.
as a result of gravitino dark matter decay. We also point out that for a 
different choice of parameters of the model, consistent with our other 
requirements, it can explain the recent FERMI observations.

We also point out a novel feature of any decaying dark
matter model which decay to photons (even as a subdominant decay mode)
that they can be used to map out the dark matter density in the galaxy.

This paper is organized as follows: in sec.~II, we present the R-parity
violation model and discuss some of its implications; in sec.~III, we
consider the gravitino as the dark matter and address its decays to show
that the model naturally predicts only three leptons as the dominant decay
final states; in sec.~IV, we present our fit to the PAMELA data;
and we conclude in sec.~V.

\section{New R-parity violating model}
We extend MSSM by adding a pair of $SU(2)_L$ triplets
$\Delta$, $\bar{\Delta}$ with hypercharge $Y=\pm 2$. The $\Delta$ field
couples to
leptons generating neutrino masses when the triplets acquire small
vacuum expectation
value (vev).
We include only the new R-parity violating interaction that involves the
$\Delta$ field and no others. In the appendix, we show that this model is
radiatively stable.

The superpotential in our model consists of three parts:
$W=W_{MSSM} + \delta W + \delta W_{\not {R}}$, where $W_{MSSM}$ is the
familiar R-parity conserving MSSM superpotential; $\delta W$ includes the new
R-parity conserving terms that involve the $\Delta$ and $\bar{\Delta}$
fields and the last term is the $\not\!{R}$ term. More explicitly,
 \begin{eqnarray}
W_{MSSM} = \lambda_u Q^T i \tau_2 H_u u^c + \lambda_d Q^T i \tau_2 H_d
d^c+ \lambda_l L^T i \tau_2 H_d e^c + \mu H_u H_d\,,
\end{eqnarray}
with usual soft terms.
\begin{eqnarray}
\delta W = f L^T i \tau_2 \Delta L + \epsilon_d H_d^T i \tau_2
\Delta H_d + \epsilon_u H_u^T i \tau_2 \bar \Delta H_u + \mu_\Delta
{\rm Tr}\left( \Delta \bar \Delta \right)\,.
\end{eqnarray}
The new soft supersymmetry (SUSY) breaking terms associated
with $\delta W$  are
\begin{eqnarray}
\delta \mathcal{L}_S = f_A \widetilde L^T i \tau_2 \Delta \widetilde
L + \epsilon_{dA} H_d^T i \tau_2 \Delta H_d + \epsilon_{uA} H_u^T i
\tau_2 \bar \Delta H_u + b_\Delta {\rm Tr}\left( \Delta \bar \Delta
\right) + h.c.\,.
\end{eqnarray}
Since the terms in $\delta W$ will be responsible for small neutrino
masses, we expect the dimensionless coupling parameters in $\delta
W$ to be very small. Roughly
speaking, the corresponding soft SUSY breaking parameters are of the
order $\epsilon_{uA} \approx \epsilon_{u} M_S$, $\epsilon_{dA}
\approx \epsilon_{d} M_S$, with $M_S$ being the SUSY breaking
scale.

Let us now discuss the R-parity violating interactions in our model.
Note that $H_{u,d}$ have no lepton number whereas $\Delta$
has $L=2$; the above superpotential therefore breaks lepton number by
two units but does not break R-parity
We now add the following interaction which violates R-parity:
\begin{eqnarray}\label{aeq}
\delta W_{\not{R}}~=~ a \Delta H_d L.
\end{eqnarray}
The associated soft breaking terms is given by:
\begin{eqnarray}\label{rhoeq}
\mathcal{L}_{\not{R}} = \rho \widetilde L \Delta H_d + {\rm h.c.}
\end{eqnarray}
Thus the two R-parity breaking parameters in our model are given by
$a$ and $\rho$. We forbid the R-parity breaking terms of MSSM, i.e.
$\lambda L L e^c + \lambda' Q L d^c + \lambda'' u^c d^c d^c + \mu' L
H_u$ from appearing in the superpotential. The question of radiative
stability of this choice of R-parity breaking is discussed in the
Appendix. We will call this type II R-Parity breaking (RPBII) as
opposed to the MSSM with R-Parity breaking which we will call type I R-Parity
breaking (RPBI). It is easy to see that unlike the RPBI models,
RPBII models are quite well hidden in low energy particle physics
processes, even for R-Parity breaking couplings of $\mathcal{O}(1)$.

 We have found a symmetry which will forbid the $LLe^c$ and the
$u^cd^cd^c$ terms of MSSM while allowing the $\Delta L H_d$ term. As far
as the $QLd^c$ term is concerned, once it is set to zero, a very
small value for this coupling is induced in higher orders after
$\Delta$ field takes a vev and it does not affect our results.

As far as gauge coupling unification is concerned, if there is no new
physics in the theory above the TeV scale, the couplings do not unify.
However, with extra intermediate scale particles, one can restore
unification of couplings. We do not address this issue in this paper.

\subsection{Baryon asymmetry and neutrino mass with R-parity violation}
As noted in the introduction, the leptophilic nature of
the PAMELA data could also be explained by keeping $LLe^c$ terms of
MSSM but the coupling strength of this interaction $\lambda$ must be
below $10^{-7}$ for both baryogenesis protection\cite{campbell} as
well as PAMELA explanation. They will then lead to one loop neutrino
masses of order ${\lambda^2 m_l}/{(16\pi^2)}\sim 10^{-7}$ eV which are
much too small to explain observations.

If one uses grand unified models such as SO(10) to
generate neutrino masses via higher dimensional operators such as
${\bf 16}_m{\bf 16}_m{\bf \overline{16}}_H{\bf \overline{16}}_H$, then in this
model the R-parity breaking interactions can come from $\lambda {\bf
16}_m{\bf 16}_m{\bf 16}_m{\bf 16}_H$ type terms after B-L symmetry
is broken by the vev of $\widetilde{\nu^c}$ field in ${\bf 16}_H$. One
could then make this couplings to be of the right order, i.e.
$\lambda \langle\widetilde{\nu^c}\rangle/{M_{Pl}}\sim 10^{-7}$. The problem
here is that it also generates $u^cd^cd^c$ type R-P violating terms
with similar strength and the model runs into conflict with proton
decay. Similar situation happen in the SU(5) model i.e. the interaction
$10_m \bar{5}_m\bar{5}_m$ that generates $LLe^c$ type R-P violating term
also generates $u^cd^cd^c$ term and therefore will have the same proton
decay problem.

The only model where only $LLe^c$ and $QLd^c$ terms can be generated
without generating the $u^cd^cd^c$ term is the model based on the gauge
group $SU(2)_L\times SU(2)_R\times SU(4)_c$ where they arise from
separate higher dimensional terms- the first two from $FFF^cF^c$ and the
last one from $F^cF^cF^cF^c$ where $F=(2,1, 4)$ and $F^c=(1,2, \bar{4})$
are the representations of the gauge group that contain fermions.

Our model, on the other hand, is a type II seesaw model with all particle
masses in the TeV range.  The constraints on the parameters of such
models
such that they do not erase any pre-existing baryon asymmetry have been
discussed in~\cite{blanchet}. The discussion for our model is similar to
this paper. A combination of the $f, h_{u,d,l}$
with either $\epsilon_{u,d}$ or $a$ will lead to violation of lepton
number, $L$. Therefore if the strengths of the $L$-violating
couplings are such that the processes caused by them are in
equilibrium in the early universe, it will erase any pre-exisitng
baryon asymmetry of the universe. From \cite{blanchet}
 we see that for neutrino masses in the
sub-eV range, the constraints on the triplet vev depend on the mass
of the $\Delta$ particle and for few hundred GeV mass of $\Delta$,
$v_T \leq 10^{-3}$ GeV.  We can therefore choose the parameters of the
model such that any pre-existing baryon asymmetry is not erased by them.

\subsection{Symmetry breaking and triplet vevs}
After electroweak symmetry is broken by the doublet Higgs fields
$\langle H_{u,d}\rangle=v_{u,d}\neq 0$,
$\Delta$ acquires an induced vev, $v_{\Delta}$.
The magnitude of triplet vev's can be estimated by minimizing the
potential of the theory to be:
\begin{eqnarray}
\langle \Delta \rangle = \left(\begin{array}{cc} 0 & 0 \\ v_T & 0
\end{array}\right), \ \  v_T = - \frac{1}{m_0^2} \left( \epsilon_{d
A} v_d^2 + \epsilon_u^* \mu_\Delta v_u^2 + \epsilon_d \mu^* v_u v_d
\right)\,, \nonumber \\
\langle \bar \Delta \rangle = \left(\begin{array}{cc} 0 & \bar{v}_{T}
\\
0 & 0 \end{array}\right), \ \  \bar{v}_{T} = - \frac{1}{m_0^2}
\left( \epsilon_{u A} v_u^2 + \epsilon_d^* \mu_\Delta v_d^2 +
\epsilon_u \mu^* v_u v_d \right)\,,
\end{eqnarray}
where $m_0$ is the
typical SUSY breaking mass and all other parameters are defined in
 the text. Rough order of magnitude of $v_{T}$ is $\sim \epsilon_{u,d}
v^2_{wk}/{M_{S}}$. If $v_T\leq $ MeV,
$\epsilon_{u,d}\leq 10^{-5}$. The triplet vev of an MeV is an input into
our model. It could arise from tadpole terms as in the usual type II
seesaw models i.e. by minimization of the terms $M^2_\Delta
\Delta^\dagger\Delta+\lambda m_{3/2}\Delta H_u H_u$ by choosing $\lambda$
of order $10^{-3}$.

\subsection{Neutrino mass and constraints from lepton flavor violation}
Neutrino masse in our model are given by type II seesaw\cite{type2},
\begin{eqnarray}\label{neutrino}
m_\nu = 2 f v_T \approx {0.1 \rm eV}\,,
\end{eqnarray}
then implies that if $v_T\leq$ MeV, $f\geq 10^{-7}$.  For $v_T\neq 0$,
there is a $\nu-\tilde{H_d}$ of magnitude $a v_T$ induced which via a
seesaw-like formula give an additional contribution to neutrino mass
$\sim \frac{(av_T)^2}{M_{SUSY}}$. For the choice of parameters in our
model, this contribution to neutrino mass is $\leq 10^{-9}$ eV and is
thus negligible.

We now turn to
 the lepton flavor violation constraints e.g. $\mu\to 3 e$ on the
couplings $f$.
In a generic triplet models of this type,
$\mu\to 3 e$ decay can arise via the exchange of the doubly charged
component of the triplet $\Delta$. Present upper limits on this process
restrict the values of the triplet coupling as follows:
$f_{11}f_{12}\leq 10^{-6}$. We will keep the $f$ coupling in the range
$10^{-7}\leq f \leq 10^{-3}$. We will see later that for natural values
of  parameters of the model, $f$ will be closer to $10^{-4}-10^{-3}$.

It is also worth pointing out that in this model after standard model
symmetry breaking, the sneutrino field will have a vev; but this vev has
a magnitude $\langle \tilde{\nu}\rangle\sim av_T\sim 10^{-11}$ GeV
which is much too small to affect any low energy leptonic physics e.g.
neutrino mass.

Furthermore, in our model, we do not have a massless Majoron
because we have explicit lepton number violation by the terms
$\epsilon_{u,d}$. Also since the triplet masses are in the $10^2$ GeV
range, there is no new contribution to the Z-width.

\section{Gravitino decay and lifetime}
To discuss dark matter and its application to PAMELA, we choose gravitino
as the lightest supersymmetric
particle (LSP) and hence the candidate for dark matter. If the theory
conserved R-parity, the gravitino would have been stable. However,
since our model has R-parity
breaking, it is unstable and as we will see below, it will have a
long lifetime so that it can be a viable dark matter of the universe. To
estimate the gravitino life time, let us look at its various decay modes
and the total width.

The gravitino can have both two and three body decays. First we will show
that the two body decay rate is very small for the choice of parameters
in the model, making the three body decays dominant. To see this, we
note that in this model at the tree level after electroweak symmetry
breaking, $\Delta-\widetilde{\ell}$ appears,
while the other
mixings such as $\widetilde{W}^- -\ell^-$ and $\widetilde{Z} -\nu$
are severely suppressed by ${\mathcal O}(\epsilon/a)\ll 1/(16\pi^2)$
comparing with $\Delta-\widetilde{\ell}$.
The {$\Delta-\widetilde{\ell}$ mixing gets contributions from
both supersymmetric as well as SUSY breaking parts and is given by
\begin{eqnarray}
U_{\widetilde{e}\Delta}\simeq \frac{(\rho
+a\mu)v_{wk}}{m^2_{\widetilde{e}}-m^2_{\Delta}}\;.
\end{eqnarray}

\begin{figure}[hbt]
\begin{center}
\includegraphics[width=0.38\textwidth]{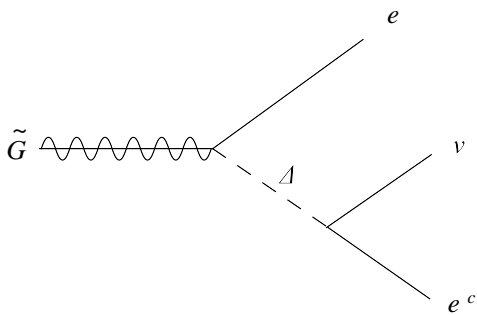}
\caption{The Feynman diagram for gravitino decays to three leptons
through a triplet Higgs.}
\label{decay}
\end{center}
\end{figure}
The three body tree level decay of the gravitino arises from the Feynman
 diagram at Fig.~\ref{decay} and the decay width for $\widetilde{G}\to
e^-e^+\nu_i$ is given by
\begin{eqnarray}
\Gamma_{\rm e^+e^-\nu_i} = \frac{|U_{\widetilde e \Delta}|^2
|f_{ei}|^2}{192\pi^3} \frac{m_{\widetilde{G}}^3}{8 M_{\rm pl}^2}F(x)\,,
\end{eqnarray}
where $x={m^2_\Delta}/{m^2_{\widetilde {G}}}$ and the function
$F(x)$ is given as
$F(x)=(2x-1)(30x^2-66x+37)/12+(x-1)^3(5x-1)\ln[(x-1)/x]$. For
$m_\Delta >m_{\widetilde{G}}$, the maximum value of $F(x)$ is about
0.04.

 For gravitino as the dark matter, we choose its lifetime to be $\sim
10^{26}$ sec,
which corresponds to the three lepton decay width
$\Gamma \sim 10^{-50}-10^{-51}$ GeV. For $\widetilde{G}$ mass
to be $\sim 300$ GeV,
 and $m_\Delta > m_{\widetilde{G}}$, this requires $|U_{\widetilde{e}\Delta}
f| \sim 10^{-8}$.

Let us now turn to the two body decays of type $\widetilde{G}\to
\nu+\gamma, \ell+W, \nu+Z$ etc. These will arise from R-parity
violating mixings between $\widetilde{\gamma}-\nu$,
$\widetilde{W}-\ell$, $\widetilde{Z}-\nu$ etc.
In our model, these mixings are
either absent or severely suppressed by
 ${\mathcal O}(\epsilon/a)\ll 1/(16\pi^2)$ at tree level. At loop level,
they appear in similar
 way as shown in Fig.~\ref{oneloop}.
\begin{figure}[hbt]
\begin{center}
\includegraphics[width=14cm]{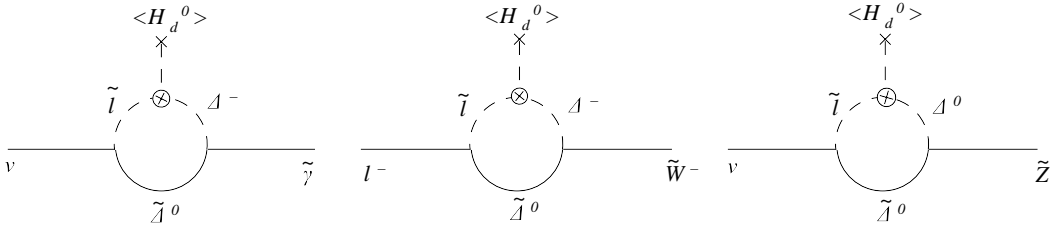}
 \caption{The one-loop contributions to the R-parity violating mixings
between $\widetilde{\gamma}-\nu$, $\widetilde{W}-\ell$,
$\widetilde{Z}-\nu$.} \label{oneloop}
\end{center}
\end{figure}
To give a typical estimate, we find for $\widetilde{\gamma}-\nu$
\begin{eqnarray}
U_{\widetilde{\gamma}-\nu}\sim
\frac{afev_d\mu}{16\pi^2M_{\widetilde{\Delta}}M_{\widetilde{\gamma}}}\,.
\end{eqnarray}
The two body decay width of the gravitino is given by:
 \begin{eqnarray}
\Gamma_{\rm two\ body} =
\frac{|U_{\widetilde{\gamma}-\nu}|^2m^3_{\widetilde{G}}}{128\pi
M^2_{Pl}}\,.
\end{eqnarray}

The ratio of two to three body decay rates can then
be given by:
 \begin{eqnarray}
\frac{\Gamma_{\rm two\ body}}{\Gamma_{\rm three\ body}} =
 \frac{|U_{\widetilde\gamma -\nu}|^2 16\pi^2}{|fU_{\widetilde{e}\Delta}|^2
4F(m^2_\Delta/m^2_{\widetilde G})/3}\,.
\end{eqnarray}
For $m_{\widetilde\Delta}\gg m_{\Delta} \simeq \ m_{\widetilde l}$,
there is a
 parameter range where the three body decay dominates. As an
example, we choose $f\sim 10^{-3}$ and  $a\sim 10^{-5}$, we find
that $|U_{\nu-\widetilde{\gamma}}|\sim 10^{-12}-10^{-13}$ or so. To
get $|fU_{\widetilde{e}\Delta}| \simeq 10^{-8}$ then implies that we
have to choose $m_{\widetilde{e}}\simeq m_\Delta$ to about 10\%
accuracy.

\section{Diffusion and PAMELA observations}

 In this section we study the positron signals from the gravitino decay.
 In this model, the gravitino dominantly decays to three body leptonic
states that
involve electrons, muons as well as tau's with triplet Higgs
mediated. To see this note that the
parameter $a$ and $\rho$ in Eq.~\ref{aeq} and~\ref{rhoeq} have flavor index. For
simplicity we could choose this to be along the electron flavor
direction. For the case of degenerate neutrinos, then, the gravitino
will have the following final states:
\begin{eqnarray}
\widetilde{G}\to e^+(\nu_e e^-, \nu_\mu \mu^-, \nu_{\tau}\tau^-)
\end{eqnarray}
In our fit we take all these modes into account.
 The positron flux
from the decay of dark matter in the halo can be obtained by solving
the steady propagation equation~\cite{Moskalenko:1999sb,
Baltz:1998xv}:
\begin{equation}
\nabla\cdot\left( K(E, \vec{x}) \nabla f_{e^{+}}\right)
+ \frac{\partial}{\partial E} \left(b(E, \vec{x}) f_{e^{+}}\right)
+Q(E, \vec{x})=0\;,
\end{equation}
 where $f_{e^{+}}$ is the number density of positron per unit energy,
$K(E, \vec{x})$ is the diffusion coefficient,
 $b(E, \vec{x})$ is the rate of energy loss and is assumed to be
spatially constant as
 $b(E)\approx 10^{-16} (E/1\textrm{GeV})^{2} \textrm{sec}^{-1}$. The
source term $Q(E, \vec{x})$ is given by
\begin{equation}
 Q(E, \vec{x}) =\frac{\rho(\vec{x})}{m_{\widetilde{G}}
\tau_{\widetilde{G}}}\frac{dN_{e^{+}}}{dE}\,,
\end{equation}
combining the distribution profile of dark matter and the energy
spectrum of positron from gravitino decay.
The solution of the transport equation at the Solar system can be expressed
by the convention~\cite{Ibarra:2008qg}
\begin{equation}
f_{e^{+}}(E)=\frac{1}{m_{\widetilde{G}} \tau_{\widetilde{G}}}
\int^{E_{max}}_{0} dE^{\prime}
G(E,E^{\prime})\frac{dN_{e^{+}}}{dE^\prime}\,,
\end{equation}
where the Green's function is well approximately given by
\begin{equation}
 G(E,E^{\prime}) \simeq
 \frac{10^{16}}{E^{2}}e^{a+b(E^{\delta-1}-{E^{\prime}}^{\delta-1})}
\theta(E^{\prime}-E) \;\textrm{cm}^{-3}\textrm{sec}\,,
\end{equation}
where the energy is in units of GeV and we adopt parameters
 $a=-1.0203, b=-1.4493, \delta=0.70$ with assuming the NFW profile and
the MED diffusion
model~\cite{Ibarra:2008qg}.

The positron flux from gravitino decay can then be obtained from
\begin{equation}
\Phi^{prim}_{e^{+}}(E)=\frac{c}{4\pi}f_{e^{+}}(E)= \frac{c}{4 \pi
m_{\widetilde{G}} \tau_{\widetilde{G}}} \int^{E_{max}}_{0}
dE^{\prime} G(E,E^{\prime})\frac{dN_{e^{+}}}{dE^\prime}\,.
\end{equation}

\begin{figure}[bht]
\begin{center}
\includegraphics[width=0.50 \textwidth]{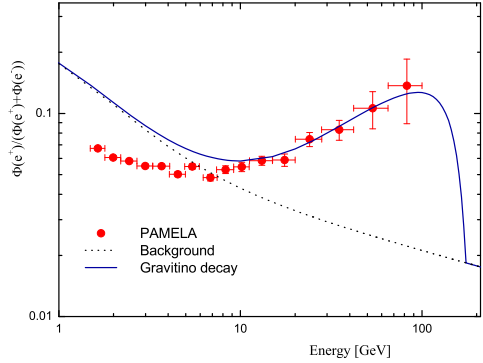}
  \caption{Fraction of positron flux as a function of energy with the
contributions from gravitino decay.}\label{pamela}
\end{center}
\end{figure}
To get the positron flux fraction, one needs to know the background
fluxes of primary and secondary
electrons and secondary positrons. We use the parametrizations obtained in
~\cite{Baltz:1998xv, Moskalenko:1997gh}
with the fluxes in units of (GeV$^{-1}$ cm$^{-2}$ sec$^{-1}$ sr$^{-1}$):
\begin{eqnarray}
\Phi^{prim}_{e^{-}}(E) &=&\frac{0.16E^{-1.1}}{1+11E^{0.9}+3.2E^{2.15}}\;,
\nonumber \\
 \Phi^{sec}_{e^{-}}(E)
&=&\frac{0.7E^{0.7}}{1+110E^{1.5}+600E^{2.9}+580E^{4.2}}\;,
\nonumber \\
\Phi^{sec}_{e^{+}}(E) &=&\frac{4.5E^{0.7}}{1+650E^{2.3}+1500E^{4.2}}\;,
\end{eqnarray}
where $E$ is expressed in units of GeV. The fraction of positron flux is then given by
\begin{equation}
\frac{\Phi^{prim}_{e^{+}} + \Phi^{sec}_{e^{+}}}{\Phi^{prim}_{e^{+}} + \Phi^{sec}_{e^{+}}+k \Phi^{prim}_{e^{-}} + \Phi^{sec}_{e^{-}}}\,,
\end{equation}
 where $k=0.88$ is a free parameter used to match the data when no primary
source of $e^{+}$ flux~\cite{Baltz:1998xv, Baltz:2001ir}.

For simplicity, we consider the degenerate neutrino mass spectrum,
in which case the triplet Higgs mainly couples to lepton pairs with
same flavor. We analytically calculate the electron, positron energy
spectrums for $\widetilde{G}\to \nu e^+e^-$ as well as their boosted
spectrums from $\mu^+\mu^- (\tau^+\tau^-)$ cascade decays. To fit
the PAMELA's data, as an example, we take $m_{\widetilde{G}}=350$
GeV, $m_{\Delta}=700$ GeV,  $|f U_{\widetilde{\ell}
\Delta}|=2.5\times 10^{-8}$, therefore the lifetime of gravitino is
about $2.1 \times 10^{26}$ sec and the fitting result is shown in
Fig.~\ref{pamela}. 

Let us now briefly comment on the recent Fermi-LAT observations~\cite{Abdo:2009zk}.
The Fermi observation shows a less pronounced $e^-+e^+$ excess above the
background in the ATIC energy range of above 100 GeV. We believe that
one can get a fit to the Fermi data if we use normal neutrino mass hierarchy
and suppress the RPV coupling involving electron superfield. The point is
that in this case, the final electrons arise mostly from the decay of the
final state muon and tau coupling to the $\Delta$ field and therefore
have a much softer spectrum.
 Without suppressing the RPV coupling involving electron superfield, one 
can choose 
 the parameters as $m_{\widetilde{G}}=3$ TeV, $m_{\Delta}=2.9$ TeV,  
 $|U_{\widetilde{\ell}
 \Delta}|=3.5\times 10^{-9}$, therefore $\tau_{\tilde{G}}=0.42 \times 
 10^{26}$ sec. In
 this case the gravitino decays to leptons and on-shell $\Delta$'s which 
will mainly produce  
 muons and tauons (then cascade decay to electrons) by choosing the 
normal neutrino mass hierarchy.
 The positron fraction and electron-positron spectrum with the 
contributions from gravitino decay is 
 given in Fig.~\ref{onshell}, where $k=0.72$ is used to normalize the 
background. We see that for this choice of parameters, both PAMELA and 
current FERMI data can be fitted well.
\begin{figure}[htb]
\begin{center}
\includegraphics[width=0.450 \textwidth]{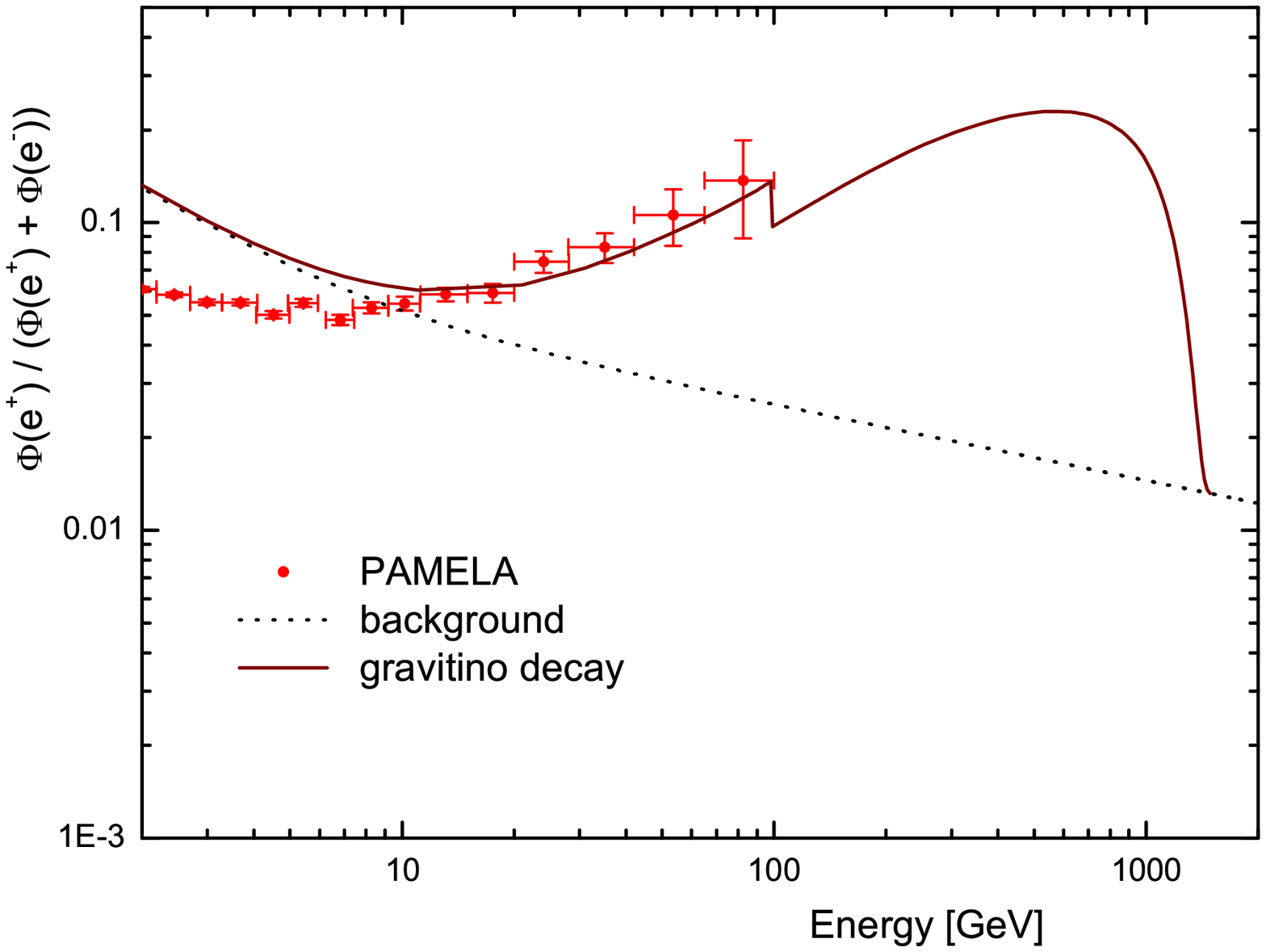}
\includegraphics[width=0.450 \textwidth]{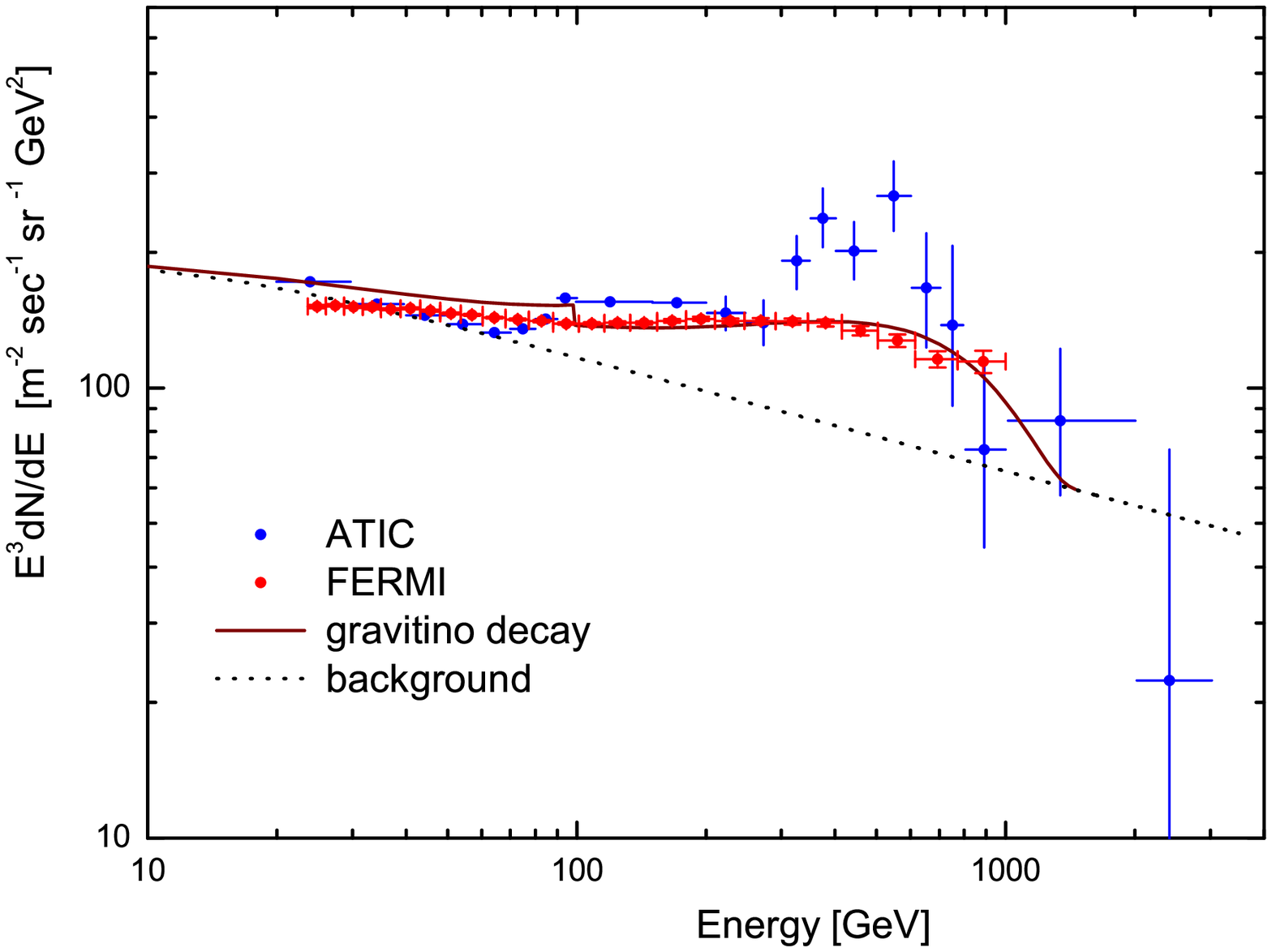}
  \caption{(left) Fraction of positron  flux as a function of energy and 
(right) the electron-positron spectrum  fit after Fermi-LAT data with the 
contributions from gravitino decay.}\label{onshell}
\end{center}
\end{figure}

\section{Conclusion}
       In conclusion, in this brief note we have proposed a
particle physics interpretation of PAMELA positron excess in terms of a
new class of R-parity violating models (RPBII) which is related to the
neutrino mass via type II seesaw mechanism. This provides a natural
explanation of why the PAMELA excess is only in positrons and and not in
hadrons. For the case of normal neutrino mass hierarchy and a 
different choice of gravitino and Delta field masses, the model can 
describe also the new FERMI data. This model also provides an explanation 
of both small neutrino
masses without invoking any physics above the TeV scale while at the same
time making sure that any pre-existing baryon asymmetry of the universe
is not erased by the R-parity violating interactions. This class of
R-parity breaking models turns out to remain very well hidden from low
energy experimental probes unlike the MSSM R-Parity breaking models.

An interesting possibility for decaying dark matter with a photon in the
final state is the opportunity to measure the dark matter density in the
halo using the intensity of the gamma rays from different directions in
satellites. We will elaborate on this idea in a future publication.

\section*{Acknowledgments}
This work was partially supported by the U. S. Department of Energy via grant DE-FG02-
93ER-40762. The works of RNM is supported by the NSF grant PHY-0652363. The work of
S. N. is partially also supported by this grant and the Maryland Center
for Fundamental Physics. Y. Z. acknowledges the
hospitality and support from the TQHN group at University of Maryland and a partial support
from NSFC grants 10421503 and 10625521.

\renewcommand{\theequation}{A\arabic{equation}}
\setcounter{equation}{0} 

\section*{Appendix: Radiative stability of the vacuum}

In this appendix, we address the issue of whether the smallness of
$\epsilon$ or $\epsilon_A$ which characterize the lepton number
violating terms are stable under radiative corrections.
>From the discussion of neutrino
mass, we know from Eq.~(\ref{neutrino}) that $\epsilon \approx
10^{-12}/f$, while in order to explain PAMELA anomaly, we demand $a
\approx 10^{-8}/f$. To suppress gravitino two-body decay, we need
to stabilize the hierarchy to ensure $\epsilon \ll
{a}/{(16\pi^2)}$.

For this purpose, note first that radiative correction to
$\epsilon$ is safe because it is a superpotential parameter
and is therefore only multiplicatively renormalized due to the
non-renormalization theorem of supersymmetry. However
 the soft term $\epsilon_A$ is not multiplicatively renormalized and there
is a contribution which is not proportional to itself, as
shown in the following figure.

\begin{figure}[hbt]
\begin{center}
\includegraphics[width=7cm]{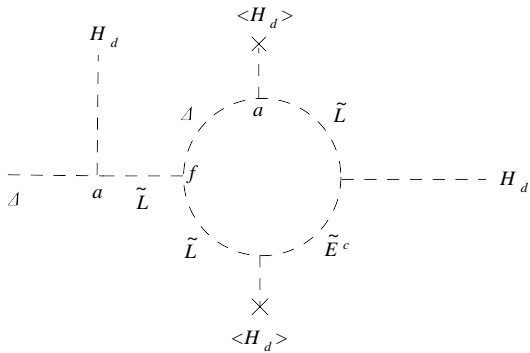}
        \caption{The radiative contribution to $H_d \Delta H_d$.}
\end{center}\label{fig4}
\end{figure}

We show what the renormalization of $\epsilon_A$ would be by a
symmetry argument. First, we are considering $H_d \Delta H_d$ soft
term, so the radiative correction must be proportional to one of the
soft parameters. Second, $H_d \Delta H_d$ violates lepton number by
2 units, so the radiative correction, if not proportional to itself,
must include two $\Delta L=1$ processes, i.e. $\propto a^2$.

Third, we can extend and restore the Peccei-Quinn (PQ) symmetry by
assigning charges to fields and
spurion's parameters. In
MSSM, the PQ charges are $Y_L = Y_{E^c} = Y_{Q} = Y_{U^c} =
Y_{D^c}=1$ and $Y_{H_u} = Y_{H_d} = -2$. The PQ symmetry is restored
if the parameters also carry charges as $Y_\mu = Y_b = 4$. Now we
extend it and set the PQ charges for triplet Higgs as $Y_{\Delta}=
Y_{\bar\Delta} = x$. From the $fL\Delta L$ term, we need assign $Y_f
= -(x+2)$. In a similar way, $Y_{\mu_\Delta} =-2x$, $Y_{a} = Y_\rho
= 1-x, Y_\epsilon = Y_{\epsilon_A} = 4-x$ demanded by the terms
$\mu_\Delta {\rm Tr}(\Delta \bar \Delta), L\Delta H_d$ and $H_d
\Delta H_d$. From the second
comment, to renormalize $\epsilon_A$ whose charge is $4-x$, we must
need to have two $a$'s, which totally contribute $2(1-x)$ charge.
Therefore it is still short of $x+2$, which is exactly the charge of
$f^*$ without other choices.

Therefore, one can conclude that the additive renormalization of
the parameter $\epsilon_A$ must take the following form
\begin{eqnarray}
\delta \left( \frac{\epsilon_A}{v_{wk}} \right) \propto
\frac{1}{16\pi^2} a^2 f^*\,,
\end{eqnarray}
which is ${10^{-18}}/{f} \ll
\epsilon_A/{v_{wk}}\approx \epsilon$. So the smallness of $\epsilon_{A}$ is
stable under radiative corrections.

We also note that the new R-parity breaking superpotential
\begin{eqnarray}
W_{\not{R}} = a L \Delta H_d
\end{eqnarray}
introduces soft R-parity breaking terms of usual MSSM variety from Fig.~\ref{rparity}.
This term will generate the usual R-parity violating MSSM terms through radiative corrections
(except the $\lambda''$ term). Their strengths are, however, very weak
and do not lead to any observable effects.

\begin{figure}[hbt]
\begin{center}
\includegraphics[width=14cm]{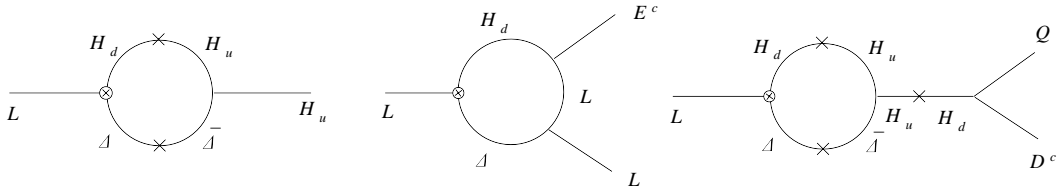}
\caption{This diagram generates the $\widetilde L-H_u$ mixing as
well as
 other soft R-parity breaking terms of MSSM.}\label{rparity}
\end{center}
\end{figure}


\begin{thebibliography}{99}

\bibitem{pamela} O.~Adriani {\it et al.}  [PAMELA Collaboration],
arXiv:0810.4995 [astro-ph]; Phys.\ Rev.\ Lett.\  {\bf 102}, 051101
(2009).

\bibitem{heat}  S.~W.~Barwick {\it et al.}  [HEAT Collaboration],
  Astrophys.\ J.\  {\bf 482}, L191 (1997).

\bibitem{ams}  M.~Aguilar {\it et al.}  [AMS-01 Collaboration],
  Phys.\ Lett.\  B {\bf 646}, 145 (2007).

\bibitem{astro}  D.~Hooper, P.~Blasi and P.~D.~Serpico,
  JCAP {\bf 0901}, 025 (2009).
  H.~Yuksel, M.~D.~Kistler and T.~Stanev,
  arXiv:0810.2784 [astro-ph].


\bibitem{ann}
  L.~Bergstrom, T.~Bringmann and J.~Edsjo,
  Phys.\ Rev.\  D {\bf 78}, 103520 (2008)
  [arXiv:0808.3725 [astro-ph]].
  M.~Cirelli and A.~Strumia,
  arXiv:0808.3867 [astro-ph].
  V.~Barger, W.~Y.~Keung, D.~Marfatia and G.~Shaughnessy,
  Phys.\ Lett.\  B {\bf 672}, 141 (2009)
  [arXiv:0809.0162 [hep-ph]].
  J.~H.~Huh, J.~E.~Kim and B.~Kyae,
  arXiv:0809.2601 [hep-ph].
  M.~Cirelli, M.~Kadastik, M.~Raidal and A.~Strumia,
  arXiv:0809.2409 [hep-ph];
  I.~Cholis, D.~P.~Finkbeiner, L.~Goodenough and N.~Weiner,
  arXiv:0810.5344 [astro-ph];
  Y.~Nomura and J.~Thaler,
  arXiv:0810.5397 [hep-ph];
  A.~E.~Nelson and C.~Spitzer,
  arXiv:0810.5167 [hep-ph];
  R.~Harnik and G.~D.~Kribs,
  arXiv:0810.5557 [hep-ph];
  D.~Feldman, Z.~Liu and P.~Nath,
  arXiv:0810.5762 [hep-ph];
  Y.~Bai and Z.~Han,
  arXiv:0811.0387 [hep-ph];
  P.~J.~Fox and E.~Poppitz,
  arXiv:0811.0399 [hep-ph];
  I.~Cholis, G.~Dobler, D.~P.~Finkbeiner, L.~Goodenough and N.~Weiner,
  arXiv:0811.3641 [astro-ph].
  G.~Bertone, M.~Cirelli, A.~Strumia and M.~Taoso,
  arXiv:0811.3744 [astro-ph].
  D.~Hooper, A.~Stebbins and K.~M.~Zurek,
  arXiv:0812.3202 [hep-ph].
  K.~J.~Bae, J.~H.~Huh, J.~E.~Kim, B.~Kyae and R.~D.~Viollier,
  arXiv:0812.3511 [hep-ph].
  L.~Bergstrom, G.~Bertone, T.~Bringmann, J.~Edsjo and M.~Taoso,
  arXiv:0812.3895 [astro-ph].
  P.~Grajek, G.~Kane, D.~Phalen, A.~Pierce and S.~Watson,
  arXiv:0812.4555 [hep-ph].
  S.~C.~Park and J.~Shu,
  arXiv:0901.0720 [hep-ph].
  I.~Gogoladze, R.~Khalid, Q.~Shafi and H.~Yuksel,
  arXiv:0901.0923 [hep-ph].
  S.~Khalil, H.~S.~Lee and E.~Ma,
  arXiv:0901.0981 [hep-ph].
  Q.~H.~Cao, E.~Ma and G.~Shaughnessy,
  arXiv:0901.1334 [hep-ph].
  J.~Hisano, M.~Kawasaki, K.~Kohri, T.~Moroi and K.~Nakayama,
  arXiv:0901.3582 [hep-ph].
  F.~Chen, J.~M.~Cline and A.~R.~Frey,
  arXiv:0901.4327 [hep-ph].
  H.~S.~Goh, L.~J.~Hall and P.~Kumar,
  arXiv:0902.0814 [hep-ph].
  R.~Allahverdi, B.~Dutta, K.~Richardson-McDaniel and Y.~Santoso,
  arXiv:0902.3463 [hep-ph].
  K.~Cheung, P.~Y.~Tseng and T.~C.~Yuan,
  arXiv:0902.4035 [hep-ph].
  D.~P.~Finkbeiner, T.~Slatyer, N.~Weiner and I.~Yavin,
  arXiv:0903.1037 [hep-ph].
  C.~R.~Chen, M.~M.~Nojiri, S.~C.~Park, J.~Shu and M.~Takeuchi,
  arXiv:0903.1971 [hep-ph].




\bibitem{decay}
  C.~R.~Chen, F.~Takahashi and T.~T.~Yanagida,
  Phys.\ Lett.\  B {\bf 671}, 71 (2009)
  [arXiv:0809.0792 [hep-ph]].
  I.~Cholis, D.~P.~Finkbeiner, L.~Goodenough and N.~Weiner,
  arXiv:0810.5344 [astro-ph].
A. Ibarra and D.~Tran,
  arXiv:0811.1555 [hep-ph]; E.~Nardi, F.~Sannino and A.~Strumia,
  JCAP {\bf 0901}, 043 (2009)
  [arXiv:0811.4153 [hep-ph]];
  P.~f.~Yin, Q.~Yuan, J.~Liu, J.~Zhang, X.~j.~Bi and S.~h.~Zhu,
  Phys.\ Rev.\  D {\bf 79}, 023512 (2009)
  [arXiv:0811.0176 [hep-ph]].
  K.~Ishiwata, S.~Matsumoto and T.~Moroi,
  arXiv:0811.0250 [hep-ph];
  C.~R.~Chen, F.~Takahashi and T.~T.~Yanagida,
  arXiv:0811.0477 [hep-ph];
  K.~Hamaguchi, E.~Nakamura, S.~Shirai and T.~T.~Yanagida,
  arXiv:0811.0737 [hep-ph];
  E.~Ponton and L.~Randall,
  arXiv:0811.1029 [hep-ph];
  A.~Ibarra and D.~Tran,
  arXiv:0811.1555 [hep-ph];
  C.~R.~Chen, M.~M.~Nojiri, F.~Takahashi and T.~T.~Yanagida,
  arXiv:0811.3357 [astro-ph];
  E.~Nardi, F.~Sannino and A.~Strumia,
  JCAP {\bf 0901}, 043 (2009)
  [arXiv:0811.4153 [hep-ph]].
  M.~Pospelov and M.~Trott,
  arXiv:0812.0432 [hep-ph].
  A.~Arvanitaki, S.~Dimopoulos, S.~Dubovsky, P.~W.~Graham, R.~Harnik
  and S.~Rajendran,
  arXiv:0812.2075 [hep-ph];
  K.~Hamaguchi, S.~Shirai and T.~T.~Yanagida,
  arXiv:0812.2374 [hep-ph];
  I.~Gogoladze, R.~Khalid, Q.~Shafi and H.~Yuksel,
  arXiv:0901.0923 [hep-ph];
    X. J. Bi, P. H. Gu, T. Li and X. Zhang, arXiv:0901.0176;
  K.~Hamaguchi, F.~Takahashi and T.~T.~Yanagida,
  arXiv:0901.2168 [hep-ph].
  C.~H.~Chen, C.~Q.~Geng and D.~V.~Zhuridov,
  arXiv:0901.2681 [hep-ph].
  X.~Chen,
  arXiv:0902.0008 [hep-ph].
  K.~J.~Bae and B.~Kyae,
  arXiv:0902.3578 [hep-ph].
  K.~Ishiwata, S.~Matsumoto and T.~Moroi,
  arXiv:0903.0242 [hep-ph].
  X.~J.~Bi, X.~G.~He and Q.~Yuan,
  arXiv:0903.0122 [hep-ph].
For an early decaying dark matter model, see K.~S.~Babu, D.~Eichler and
R.~N.~Mohapatra,
  Phys.\ Lett.\  B {\bf 226}, 347 (1989);


\bibitem{Navarro:1995iw}
  J.~F.~Navarro, C.~S.~Frenk and S.~D.~M.~White,
  Astrophys.\ J.\  {\bf 462}, 563 (1996)
  [arXiv:astro-ph/9508025].

\bibitem{new} M.~Cirelli, A.~Strumia and M.~Tamburini,
  Nucl.\ Phys.\  B {\bf 787}, 152 (2007)
  [arXiv:0706.4071 [hep-ph]];  N.~Arkani-Hamed, D.~P.~Finkbeiner,
T.~Slatyer and N.~Weiner,
  Phys.\ Rev.\  D {\bf 79}, 015014 (2009)

\bibitem{gravi}  W.~Buchmuller, L.~Covi, K.~Hamaguchi, A.~Ibarra and
T.~Yanagida,
  JHEP {\bf 0703}, 037 (2007);
Laura Covi, Michael Grefe, Alejandro Ibarra,
David Tran, JCAP {\bf 0901}, 029 (2009);  X.~Ji, R.~N.~Mohapatra,
S.~Nussinov and Y.~Zhang, Phys.\ Rev.\  D {\bf 78}, 075032 (2008).

\bibitem{bar}  R.~Barbier {\it et al.},
  Phys.\ Rept.\  {\bf 420}, 1 (2005)

\bibitem{Ibarra:2008qg}
  A.~Ibarra and D.~Tran,
  JCAP {\bf 0807}, 002 (2008)
  [arXiv:0804.4596 [astro-ph]].

\bibitem{type2}  W.~Konetschny and W.~Kummer,
  Phys.\ Lett.\  B {\bf 70}, 433 (1977);
R.E. Marshak, R. N. Mohapatra
Invited talk given at Orbis Scientiae, Coral Gables, Fla., Jan 14-17,
1980 (Published in the proceedings p. 277);
  T.~P.~Cheng and L.~F.~Li,
  Phys.\ Rev.\  D {\bf 22}, 2860 (1980);
  G.~Lazarides, Q.~Shafi and C.~Wetterich,
  Nucl.\ Phys.\  B {\bf 181}, 287 (1981);
  J.~Schechter and J.~W.~F.~Valle,
  Phys.\ Rev.\  D {\bf 22}, 2227 (1980);
  R.~N.~Mohapatra and G.~Senjanovic,
  Phys.\ Rev.\  D {\bf 23}, 165 (1981).


\bibitem{campbell} B. A. Campbell et al. Phys. Lett. {\bf B 256}, 457
(1991); H.
Dreiner and G. G. Ross, Nucl. Phys. {\bf B 410}, 188 (1993).

\bibitem{blanchet}   S.~Blanchet, Z.~Chacko and R.~N.~Mohapatra,
  arXiv:0812.3837 [hep-ph].


\bibitem{Moskalenko:1999sb}
  I.~V.~Moskalenko and A.~W.~Strong,
  Phys.\ Rev.\  D {\bf 60}, 063003 (1999)
  [arXiv:astro-ph/9905283].

\bibitem{Baltz:1998xv}
  E.~A.~Baltz and J.~Edsjo,
  Phys.\ Rev.\  D {\bf 59}, 023511 (1999)
  [arXiv:astro-ph/9808243].


\bibitem{Moskalenko:1997gh}
  I.~V.~Moskalenko and A.~W.~Strong,
  Astrophys.\ J.\  {\bf 493}, 694 (1998)
  [arXiv:astro-ph/9710124].

\bibitem{Baltz:2001ir}
  E.~A.~Baltz, J.~Edsjo, K.~Freese and P.~Gondolo,
  Phys.\ Rev.\  D {\bf 65}, 063511 (2002)
  [arXiv:astro-ph/0109318].

\bibitem{Abdo:2009zk}
  A.~A.~Abdo {\it et al.}  [The Fermi LAT Collaboration],
  arXiv:0905.0025 [astro-ph.HE].


 \end{thebibliography}
\end{document}